\documentclass[aps,twocolumn,showpacs]{revtex4-1}
\usepackage{graphicx,amssymb,amsmath,amsfonts}
\usepackage{MnSymbol}
\begin{document}
\title{CONVEX GEOMETRY: a travel to the limits of our
  knowledge}

\author{Bogdan Mielnik} 

\affiliation{Physics Department, Centro de
  Investigaci\'on y de Estudios
  Avanzados del IPN,
  Mexico DF, Mexico,
  Email: bogdan@fis.cinvestav.mx}

\date{February 3, 2012}

\begin{abstract}
  Our knowledge and ignorance concerning the geometry of quantum
  states are discussed.
\end{abstract}
\pacs{03.65.Aa, 02.40.Ft
}

\maketitle

\section*{Questions about the structure}

Physical theories are usually created by accumulating some fragments
of information which at the beginning do not allow to predict the
final structures. The classical mechanics was formulated by Isaac
Newton in terms of mass, force, acceleration and the three dynamical
laws. It was not immediate to see the Lagrangians, Hamilton equations
and the simplectic geometry behind. We cannot guess the reaction of
Newton if he were informed that he was just describing the classical
phase spaces defined by the simplectic manifolds\ldots

Quite similarly, Max Planck, Niels Bohr, Louis de~Broglie, Erwin
Schr\"o\-din\-ger and Werner Heisenberg could not see from the very
beginning that the physical facts which they described would be
reduced by Born's statistical interpretation to the Hilbert space
geometry (as it seems, neither Hilbert could predict that). Yet, once
accepted that the pure states of a quantum system can be represented
by vectors of a complex linear space and the expectation values are
just quadratic forms, the Hilbert spaces entered irremediably into the
quantum theories. Together appeared the ``density matrices'' as the
mathematical tools representing either pure or mixed quantum
ensembles. Their role is now so commonly accepted that its origin is
somehow lost in some petrified parts of our subconsciousness: an
obligatory element of knowledge which the best university students
(and the future specialists) learn by heart. However, is it indeed
necessary? Can indeed the interference pictures of particle beams
limit the fundamental quantum concepts to vectors in linear spaces and
``density matrices''?

\section*{Quantum logic?}

The desire to find some deeper reasons led a group of authors to
postulate the existence of an ``intrinsic logic'' of quantum
phenomena, called the \emph{quantum
  logic}~\cite{rf1,rf2,rf3}. Generalizing the classical ideas, it was
understood as the collection $Q$ of all statements (informations)
about a quantum object, possible to check by elementary ``yes-no
measurements''. Following the good traditions, $Q$ should be endowed
with \emph{implication} ($\Rightarrow$), and \emph{negation}
$a\rightarrow a'$. The implication defines the partial order in $Q$
($a\Rightarrow b$ reinterpreted as $a\leq b$), suggesting the next
axioms about the existence of the lowest upper bound $a\lor b$ (``or''
of the logic) and the greatest lower bound $a\wedge b$ (``and'' of the
logic) for any $a,b\in Q$. The ``negation'' was assumed to be
involutive, $a''\equiv a$, satisfying de Morgan law: $(a\lor b)'\equiv
a'\wedge b'$ as well as other axioms granting that $Q$ is an
orthocomplemented lattice~\cite{rf1}. Until now, the whole structure
looked quite traditional. With one exception: in contrast to the
classical measurements, the quantum ones do not commute, which
traduces itself into breaking the \emph{distributive law} $(a\lor
b)\wedge c \neq(a\wedge c)\lor(b\wedge c)$ obligatory in any classical
logic. The quantum logic was non-Boolean! An intense search for an
axiom which would generalize the distributive law, admitting both
classical and quantum measurements, in agreement with Birkhoff,
von~Neumann, Finkelstein~\cite{rf1,rf2,rf3} and thanks to the
mathematical studies of Varadarajan~\cite{rf4} convinced C.~Piron to
propose the \emph{weak modularity} as the unifying law. To some
surprise, the subsequent theorems~\cite{rf4,rf5} exhibit certain
natural completeness: the possible cases of ``quantum logic'' are
exhausted by the Boolean and Hilbertian models, or by combinations of
both. As pointed out by many authors this gives the theoretical
physicists some reasonable confidence that the formalism they develop
(with Hilbert spaces, density matrices, etc.) does not overlook
something essential, so there will be no longer need to think too much
about abstract foundations.

However, isn't this confidence a bit too scholastic? It can be noticed
that the general form of quantum theory, since a long time, is the
only element of our knowledge which does not evolve. While the
``quantization problem'' is formulated for the existing (or
hypothetical) objects of increasing dimension and flexibility (loops,
strings, gauge fields, submanifolds or pseudo-Riemannian spaces,
non-linear gravitons, etc.), the applied quantum structure is always
the same rigid Hilbertian sphere or density matrix insensitive to the
natural geometry of the ``quantized'' systems. The danger is that (in
spite of all ``spin foams'') we shall invest a lot of effort to
describe the relativistic space-times in terms of the perfectly
symmetric, ``crystalline'' forms of Hilbert spaces, like rigid bricks
covering a curved highway. Is there any other option?\ldots

\section*{Convex geometry}

The alternatives arise if one decides to describe the statistical
theories in terms of geometrical instead of logical concepts. What is
the natural geometry of the statistical theory? It should describe the
pure or mixed particle \emph{ensembles} (also ensembles of
multiparticle systems, including the mesoscopic or macroscopic
objects). Suppose that one is not interested in the total number of
the ensemble individuals, but only in their ``average
properties''. Two ensembles with the same statistical averages cannot
be distinguished by any statistical experiments: we thus say that they
define the same \emph{state}. Now consider the set $S$ of all
\emph{states} for certain physical objects. Even in absence of any
analytic description, there must exist in $S$ some simple empirical
geometry. Given any two states $x_{1},x_{2}\in S$ (corresponding to
certain ensembles $\mathcal{E}_{1}$, $\mathcal{E}_{2}$) and two numbers
$p_{1},p_{2}\geq 0$, $p_{1}+p_{2}=1$, consider a new ensemble
$\mathcal{E}$ formed by choosing randomly the objects of
$\mathcal{E}_{1}$ and $\mathcal{E}_{2}$ with probabilities $p_{1}$ and
$p_{2}$; its state, denoted $x=p_{1}x_{1}+p_{2}x_{2}$ is the
\emph{mixture} of $x_{1}$ and $x_{2}$ in proportions $p_{1},p_{2}$. If
in turn both $x_{1},x_{2}$ are mixtures of $y_{1},y_{2}\in S$, then
some more information is needed to determine the contents of $y_{1}$
and $y_{2}$ in $x$. It can be most simply expressed by representing
$S$ as a subset of a certain \emph{affine space} $\Gamma$, where
$p_{1}x_{1}+p_{2}x_{2}$ becomes a linear combination. For any two
points $x_{1},x_{2}\in S$ all mixtures $p_{1}x_{1}+p_{2}x_{2}$
($p_{1},p_{2}\geq0$, $p_{1}+p_{2}=1$) form then the straight line
interval between $x_{1}$ and $x_{2}$, contained in $S$. Hence, $S$ is
a convex set~\cite{rf6,rf7}. To describe the limiting transitions,
$\Gamma$ must possess a topology and $S$ should be closed in $\Gamma$.

\begin{figure}[htb]
  \centering
  \includegraphics[width=0.8\linewidth]{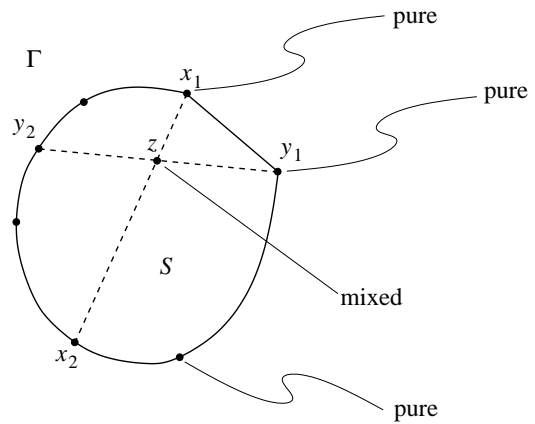}
  \caption{A convex set in 2D. Supposing that it could represent the
    states of some hypothetical ensembles, all border points except
    the straight line interval joining $x_{1}$ and $y_{1}$ would
    represent the pure states. All points in the interior are mixed
    states and do not allow a unique definition of the pure
    components. Thus, e.g., the state $z$ could be represented as a
    mixture of $x_{1}$ and $x_{2}$ or $y_{1}$ and $y_{2}$ or in any
    other way. \label{fig:1}}
\end{figure}

The information encoded in the convex structure of $S$ might seem poor:
it tells only which states are mixtures of which other states (see
Fig.~\ref{fig:1}). Yet it turns out that it contains all essential
information about both, logical and statistical aspects of quantum
theory.
\section*{Logic of properties}

The boundary of $S$ contains some special points $x$, which \emph{are
  not} untrivial combinations $p_{1}x_{1}+p_{2}x_{2}$ with
$p_{1},p_{2}> 0$ of any two different points $x_{1}\neq x_{2}$
of~$S$. These points, called \emph{extremal}, represent the physical
ensembles which \emph{are not mixtures} of different components, and
so are called \emph{pure}. All subensembles of a pure ensemble define
the same \emph{pure state} $x$, which therefore represents also the
quality of each single ensemble individual.

\begin{figure}[ht]
  \centering
  \includegraphics[width=0.6\linewidth]{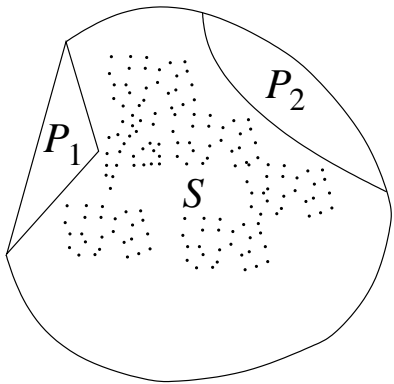}
  \caption{``Faces'' on the border of $S$ represent properties of the
    single ensemble individuals. The picture in perspective permits to
    see that $P_{1}$ and $P_{2}$ are not orthogonal.}
  \label{fig:2}
\end{figure}

The convex geometry permits to describe as well more general ensemble
properties which might be attributed to the single individuals. Note
that, in general, the property of ensemble is not shared by the
individuals (e.g., a human ensemble can contain 50\%\ of men and 50\%\
of women, but each individual, in general, has only one of these
qualities). We say that the subset $P\subset S$ defines a
\emph{property} of the single objects if: 1.~it resists mixing, i.e.,
$y_{1},y_{2}\in P$, $p_{1},p_{2}\geq 0$, $p_{1}+p_{2}=1$ $\Rightarrow$
$p_{1}y_{1}+p_{2}y_{2}\in P$ (meaning that $P$ is a convex subset of
$S$), 2.~if the property of mixture is shared by every mixture
components, i.e., $y\in P$ and $p_{1}y_{1}+p_{2}y_{2}$,
$y_{1},y_{2}\in S$, $p_{1},p_{2}> 0\Rightarrow y_{1},y_{2}\in P$. The
subset $P\subset S$ which satisfies 1.~and 2. is called a \emph{face}
of $S$. The whole $S$ and the empty set $\emptyset$ are the improper
faces: all other are plane fragments of various dimensionalities on
the boundary of $S$ (See Fig.~\ref{fig:2}). In particular, each
extremal points is a one point face. In what follows, we shall be most
interested in the topologically \emph{closed} faces of $S$
representing the ``continuous properties'' of the ensemble
objects. Further on by \emph{faces} we shall mean \emph{closed
  faces}. Their whole family $\mathcal{P}$ admits a partial ordering
$\leq$ identical with the set theoretical inclusion: the relation
$P_{1}\leq P_{2} \Leftrightarrow P_{1}\subset P_{2}$ means that the
property $P_{1}$ is \emph{more restrictive} than $P_{2}$, or $P_{1}$
\emph{implies} $P_{2}$. As easily seen, the intersection of any family
of faces is again a face of $S$. Hence, for any two faces
$P_{1},P_{2}\subset S$ there exists also their \emph{smallest upper
  bound}, or \emph{union} $P_{1}\lor P_{2}$, defined as the
intersection of all faces containing both $P_{1}$ and $P_{2}$. The set
$\mathcal{P}$ with the partial order $\leq$ (i.e., implication) and
operations $\lor$, $\wedge$ is thus a complete lattice generalizing
the ``quantum logic'' of the orthodox quantum mechanics: it might be
called the \emph{logic of properties}. We shall see that it does not
necessarily include \emph{negation}, but it admits a natural concept
of \emph{orthogonality}~\cite{rf6,rf8}.

\section*{Counters}

A natural counterpart of quantum ensembles are the measuring devices
and the simplest such devices are \emph{particle counters}. By a
\emph{counter} we shall understand any macroscopic body sensitive
(either perfectly or partly) to the presence of quanta. Our assumption
is also, that each counter reacts only to the properties of each
single ensemble individual, without depending on the the rest. In
mathematical terms, each counter can be described by a certain
functional $\phi:S\rightarrow[0,1]$. whose values $\phi x$ for any
$x\in S$ mean the fraction of particles in the state $x$ detected by
the counter $\phi$. If $\phi x=1$, then the counter $\phi$ detects
perfectly all $x$-particles, if $0<\phi x<1$, it overlooks a part, but
if $\phi x=0$, then $\phi$ is completely blind to the
$x$-particles. Moreover, if $\phi$ reacts only to single ensemble
individuals, then for any mixed state $x=p_{1}x_{1}+p_{2}x_{2}$ it
will detect independently both mixture components: $\phi x=p_{1}\phi
x_{1}+p_{2}\phi x_{2}$, meaning that $\phi$ is a \emph{linear
  functional} on $S$. We shall assume, that the values of counters
permit to distinguish the different points $x\in S$ and moreover, they
induce a physically meaningful topology, in which they are \emph{eo
  ipso} continuous. Each continuous, linear functional $\phi$ taking
on $S$ the values $0\leq\phi x\leq 1$ will be called
\emph{normal}. Mathematically, the counters are, therefore, the
\emph{normal functionals}. To get their geometric image, assume that
the surrounding affine space $\Gamma \supset S$ is spanned by
$S$. Hence, every linear functional $\phi$ on $S$ defines a unique
linear functional on $\Gamma$ which will be denoted by the same symbol
$\phi$. If $\phi\equiv\text{const.}$ on $S$, then
$\phi\equiv\text{const.}$ on $\Gamma$. If not, then the equations
$\phi x=c$, ($c\in\mathbb{R}$) split $\Gamma$ into a continuum of
parallel hyperplanes on which $\phi$ accepts distinct constant
values. Due to the linearity, $\phi$ is completely defined by the pair
of hyperplanes on which it takes the values~0 and~1. If $\phi$ is
normal, $S$ is contained in the closed belt of space between both
planes. The question arises, how ample is the set of physical
counters? Since no restrictions are evident, we shall assume that each
normal functional represents a particle counter which at least in
principle can be constructed. All distinct ways of counting particles
can be thus read from the convex geometry of $S$~\cite{rf8}. They turn
out closely related with the collection of hyperplanes and those are
related with \emph{faces}. Indeed, the hyperplanes $\phi=1$ and
$\phi=0$ of a counter do not cross the interior of $S$, but can
``touch'' its boundary. As one can easily show, their common parts
with the border $\partial S$ are two ``opposite'' faces (properties)
of $S$, which awake completely different reactions of the counter:
while detecting all particles on one of them, it ignores completely
the particles on the other. Any two faces $P_{1}$, $P_{2}$, for which
there exists at least one, so sharply discriminating counter, will be
called \emph{excluding} or \emph{orthogonal} $(P_{1}\perp P_{2})$. The
``logic of properties'', therefore, is a lattice with the relations of
\emph{inclusion} $(\leq)$ and \emph{exclusion} $(\perp)$ though
without a unique ortho-\emph{complement} (since for any
$P\in\mathcal{P}$, amongst all elements orthogonal to $P$ no greatest
one must exist).

\section*{Detection ratios}

\begin{figure}[htb]
  \centering
  \includegraphics[width=0.6\linewidth]{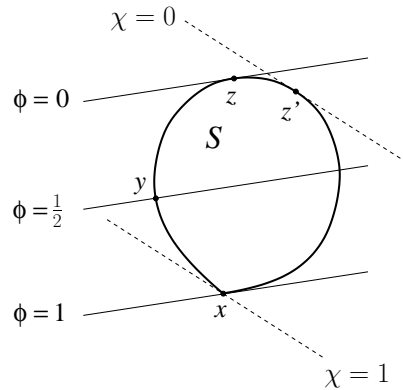}
  \caption{The \emph{could be} convex set $S$ for some hypothetical
    ensembles. The parallel support lines $\chi=1$ and $\chi=0$
    represent a counter detecting all $x$-particles, blind to
    $z'$-particles, while the lines $\phi=1$ and $\phi=0$ correspond
    to another counter, detecting all $x$-particles, but the minimal
    possible fraction $\phi y=\frac{1}{2}$ of the
    $y$-particles. Hence, the detection ratio $y:x=\frac{1}{2}$.}
  \label{fig:3}
\end{figure}

Apart from orthogonality, the next geometry element of $S$ describes
the selectivity limits of quantum measurements. Given a pair of pure
states $x,y\in S$, consider the family of all counters $\phi$
detecting unmistakeably all particles of the state $x$, i.e., $\phi
x=1$. Can they ignore completely the particles of the state $y$? In
general, the answer is negative. The following lower bound over the
counters $\phi$:
\begin{equation}
  \label{eq:1}
  y:x=\operatorname{inf}_{\phi x=1}\phi y
\end{equation}
called the ``detection ratio''~\cite{rf8}, if non-vanishing, describes
a minimal fraction of $y$-particles which must infiltrate any
experiment programmed to detect the $x$-state. The geometric character
of this quantity is defined just by convex structure of $S$, which
determines the support planes (see Fig.~\ref{fig:3}).
The information contained in~\eqref{eq:1} might be significantly
weaker if the pure state $x$ were not \emph{exposed}, i.e., determined
completely as the intersection of $S$ and at least one support
hyperplane. Such cases do not occur in the orthodox QM, but belong to
the general convex set geometry (see~\cite[Fig12]{rf9}).

\section*{The orthodox geometry}

In the orthodox theory the pure states are represented by vectors in a
complex, linear space (an inspiration from the observed interference
patterns) and all measured expectation values are real, quadratic
forms of the state vectors $\psi$ (the consequence of Born's
statistical interpretation). The mixed states are the probability
measures on the manifold of pure states (the projective Hilbert
sphere). However, since the statistical averages are no more than
quadratic forms, the ample classes of probability measures
(interpreted as the prescriptions of forming mixtures) are physically
indistinguishable. The faithful representation of the mixed states as
the equivalence classes explains the origin of the ``density
matrices''~\cite{rf8}.

The elements of the convex geometry provide also an alternative
interpretation of the unitary invariants
$\vert\langle\psi,\varphi\rangle\vert^{2}$ called currently the
``transition probabilities''. In fact if $S$ is the convex set of
density matrices in a Hilbert space and if two pure states are
represented as $x=\vert \psi\rangle\langle\psi\vert$ and  $y=\vert
\varphi\rangle\langle\varphi\vert$ ($\|\psi\|=\|\varphi\|=1$) then the
elementary lemma shows that
\begin{equation}
  \label{eq:2}
  \vert \varphi\rangle\langle\varphi\vert : \vert
  \psi\rangle\langle\psi\vert =
  \vert\langle\psi,\varphi\rangle\vert^{2}
\end{equation}
i.e., the commonly used invariant turns out the detection
ratio~\cite{rf8}, revealing an additional sense of the ``transition
probabilities''. In fact, as once noticed by Peter Bergman, the
deepest picture of a physical theory is obtained not so much by
telling what is possible, but rather by ``no go principles'', defining
what is ruled out (e.g., the equivalence principle in General
Relativity, or the uncertainty principle in QM). One such law emerges
from the identity~\eqref{eq:2}. 
\begin{figure}[htb]
  \centering
  \includegraphics[width=0.6\linewidth]{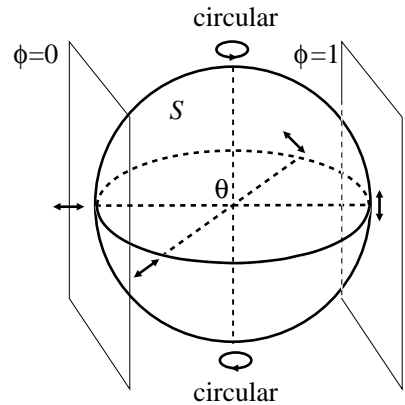}
  \caption{Multiple experiments justify the representation of the
    photon polarization states in form of the 1-qubit (Bloch) sphere
    in $\mathbb{R}^{3}$. Once fixed the image, the geometry of $S$
    determines uniquely the ``transition probabilities'' between any
    pair of pure states. On the figure: the pair of support planes
    $\phi=1$ and $\phi=0$ illustrates the maximally selective counter
    which detects all photons in the vertical polarization
    $\updownarrow$ , but none in the horizontal $\leftrightarrow$. The
    intermediate values of $\phi$ on the congruence of parallel planes
    intersecting $S$ define the transition probabilities from all
    other states to the vertical one~$\updownarrow$.}
  \label{fig:4}
\end{figure}
Indeed, $\vert\langle\psi,\varphi\rangle\vert^{2}$ not only defines
the fraction of the $\varphi$-particles accepted by the $\psi$-filter,
but also the fundamental impossibility of accepting less! Every
physical process which leads to a certain macroscopic effect for
\emph{all} $\psi$-particles, must lead to the same effect at least for
the fraction $\vert\langle\psi,\varphi\rangle\vert^{2}$ of
$\varphi$-particles. The purely geometric nature of this law,
independent of any analytic expression can be best illustrated by the
Bloch sphere of the photon polarization states (see Fig.~\ref{fig:4})
on which the linear polarizations occupy a great circle (the
``equator''), circular polarizations are the poles, and the remaining
surface points are the elliptic polarizations.  The interior of the
sphere collects the mixed polarizations, the center
$\theta$~representing the complete polarization chaos. The pair of
tangent planes $\phi=1$~and $\phi=0$~represents a maximally selective
counter detecting all photons in the vertical polarization and
rejecting the orthogonal one. The geometry of the sphere $S$
determines immediately the ``transition probabilities'' between any
two pure states without the need of using the analytic $\vert
\varphi\rangle\langle\varphi\vert$ representation (thus, e.g., the
detection ratio between any linear and circular polarization is 1/2).

In case of non-classical ensembles, the geometry of $S$ expresses
still more fundamental law about the indistinguishability of quantum
mixtures, the phenomenon which appears if $S$ is not a
simplex. \emph{Given a mixed ensemble of non-classical objects, one
  cannot, in general, retrospect and find out how the mixture has been
  prepared. Two mixtures composed of different collections of pure
  states can be physically indistinguishable (see also Fig.~\ref{fig:1}).}

In the Bloch sphere of polarization states (Fig.~\ref{fig:4}) the
effect is exceptionally simple for its center $\theta$ which can be
represented equivalently as a mixture of any pair of orthogonal linear
polarizations, or two opposite circular polarizations or in any other
way:
\begin{equation}
  \label{eq:3}
  \begin{split}
    \theta&\equiv\frac{1}{2}\updownarrow+\frac{1}{2}\leftrightarrow\\
    &\equiv\frac{1}{2}\neswarrow+\frac{1}{2}\nwsearrow\\
    &\equiv\frac{1}{2}\lcirclearrowleft+\frac{1}{2}\rcirclearrowright\\
    &\equiv\cdots\cdots
  \end{split}
\end{equation}
Hence, once having the mixed state $\theta$ one cannot go back and
identify its pure components: a kind of statistical paradox making it
quite difficult to check experimentally some semantic curiosities of
the existing theory, which are still waiting for a good inspiration!

\section*{Generalized geometries: are they possible?}

The structures reported here contain a certain puzzle. It is basically
not strange that the convex geometry is a language of statistical
theories. Yet, it was not expected that the structure of an arbitrary
convex set $S$ contains the equivalents of principal quantum
mechanical concepts. Their properties are distorted, but their meaning
is similar. Thus, the \emph{logic of properties} is an analogue of the
\emph{quantum logic}~\cite{rf1} and the \emph{detection ratios} are
equivalents of the orthodox ``transition probabilities''. In many
aspect the Hilbertian schemes are distinguished by their maximal
regularity and almost crystalline symmetry: to each face of $S$, (read
subspace), corresponds a unique orthocomplement, etc. Might this
resemble the relation between the Euclidean and Riemannian geometries?
If so then could it happen that in some circumstances the quantum
systems could obey the generalized convex geometry, dissenting from
the Hilbertian structure?

In the intents of finding a synthesis of the lattice (``logical'') and
probabilistic interpretations (since J.~von~Neumann~\cite{rf2}) the
statistical aspects, in general, were subordinated to the assumed
structure of the orthocomplemented lattice, and the answer of the
axiomatic approach was always the same: the quantum mechanics must be
exactly as it is. This belief turned even stronger due to the theorem
of Gleason~\cite{rf10}, as well as due to the profound and elegant
generalizations of the algebraic approaches of Gel'fand and
Naimark~\cite{rf11}, Haag and Kastler~\cite{rf12}, Pool~\cite{rf13},
Araki~\cite{rf14}, Haag~\cite{rf15}, and other authors, who never
resigned from the Hilbert space representations. Curiously, until
today, these convictions find also also a strong support in the well
known book of G.~Mackey~\cite{rf16} in which, however, the axiomatic
approach has some self-annihilating aspects: after a laborious
presentation of six axioms on quantum logic $\mathcal{L}$, the seventh
axiom tells flatly that the elements of the logic are closed vector
subspaces of a Hilbert space, thus making all previous axioms
redundant! (a short report on this school of axiomatics, see
H.~Primas~\cite[p.~211]{rf17}). Some opposition is not so
surprising\ldots

The first descriptions of QM based exclusively on the convex geometry
belong to G.~Ludwig~\cite{rf18}, though the author adopted axioms in
fact limiting the story to the orthodox scheme. The hypothesis about
the possibility of quantum mixtures obeying non-Hilbertian geometries
was formulated by the present author~\cite{rf6,rf8}, also by Davies
and Lewis~\cite{rf19}. The hypothetical geometries succeeded to awake
both positive and hostile reactions. Roger Penrose at some moment
hoped that the atypical structures might tell something about the
nonlinear graviton~\cite{rf20}, though later on he
complained~\cite{rf21} that they give a pure statistical
interpretation, without any analytical entity behind (though inversely,
the nonlinear graviton of Penrose is a pure analytical entity without
any statistical interpretation!). T.W.~Kibble and S.~Randjbar-Daemi
followed~\cite{rf8} describing the \emph{classical gravity} in
interaction with the generalized quantum structure~\cite{rf22}. Some
other authors in philosophy of physics stay firmly on the ground of
the orthodox theory. Nonetheless, they don't escape objections. While
Putnam considers the orthocomplemented structure of Hilbert spaces the
``truth of quantum mechanics''~\cite{rf23} (taking the side of
Mackey?), John Bell and Bill Hallet~\cite{rf24} adopt the generalized
design proposed in~\cite{rf6} to show the weakness of Putnam's
argument. However, the deformed geometries, if real, must occur in
some concrete physical circumstances. Where should we look for them?

As it seems, the most natural possibility is to look for nonlinear
variants of quantum mechanics. In fact, already some simple nonlinear
cases of the Schr\"odinger's equation admit nonquadratic, positive,
absolutely conservative quantities which could be used to define the
probability densities~\cite{rf8}. The quantum mechanics with
logarithmic nonlinearity permits to define consistently the reduction
of the wave packets~\cite{rf25}. Yet, as shown by Haag and
Bannier~\cite{rf26}, subsequently also in~\cite{rf27}, the nonlinear
wave equations lead to high mobility of quantum states, breaking the
quantum impossibility principles.

The most basic difficulty was noticed by N.~Gisin, who had shown that
if the linear evolution law of quantum states were amended by adding
some nonlinear operations, then the breaking of the mixture
indistinguishability would make possible to read the instantaneous
messages between parts of the entangled particle
systems~\cite{rf28,rf28a}. The simplest case would occur in a variant
of EPR experiment for the sequences of photon pairs in the singlet
polarization state
$\vert\Xi\rangle=\frac{1}{\sqrt{2}}(\vert\updownarrow\rangle
\vert\leftrightarrow\rangle -\vert\leftrightarrow\rangle
\vert\updownarrow\rangle)$ emitted in two opposite directions.
According to the present day theory the polarization measurements on
the left photons can produce at distance (due to the correlation
mechanism) any desired mixture~\eqref{eq:3} of the right photons (or
vice versa). As long as mixtures~\eqref{eq:3} are indistinguishable,
this does not transmit information. However, if the observer of the
right photon states could cause their nonlinear evolution, he could
distinguish the quantum mixtures~\eqref{eq:3}, thus reading hidden
information and reconstructing without delay the measurements
performed by his distant counterpart on the left EPR photons. So, is
the nonlinear QM impossible?

Perhaps, we should not overestimate the axiomatic approaches.  What
they usually tell is that we cannot modify just one element of the
theory, while leaving the whole rest intact. If in the last decade of
XIX century some excellent axiomaticians tried to formulate reasonable
axioms defining the space-time structure, they would prove beyond any
doubt that the space-time must be Galilean!  Yet, it is not. The
deviations (in our normal conditions) are very small, but rather
important\ldots

What can be impossible in QM, is to conserve the orthodox
representation of pure states as the ``rays'' in a complex Hilbert
space, together with the tensor product formalism, and with the
unitary background evolution, but to extend it by adding some
nonlinear evolution operations and to expect that the instantaneous
information transfers will be still blocked.  However, the whole
deduction might be already overloaded by too many axioms. If the
evolution were extended by some nonlinear operations, then in the
first place, we would loose the Hilbert space orthogonality together
with the trace rules for probabilities even without worrying about the
superluminal messages\ldots

Returning to the spin or polarization qubits, the possibilities of
generalizing the Hilbertian structures depends not so much on axioms
but rather on precise knowledge of probabilities. If indeed exactly
orthodox, then may be, the qubits can only rigidly rotate\ldots

The problems of systems traditionally described by multi- or
infinitely-dimensional Hilbert spaces are more difficult. The
questions of Hans Primas, perhaps are still waiting for a good answer:
\emph{Does quantum mechanics apply to large molecular systems?\ldots}
\emph{Why do so many stationary states not exist?}  (see~\cite[p.~11
and~12]{rf18}). Indeed, even the problem of how to create in practice
the one particle states described by arbitrary wave packets deserves
systematic studies~\cite{rf29,rf30,rf31,rf32}.

As recently noticed, the non-linear modifications of quantum dynamics
instead of just extending the techniques of the state manipulation
might introduce \emph{constrains}, with the restricted $S$ no longer
obeying the Hilbertian geometry~\cite{rf33}; an option which might be
worth exploring.

All the attempts to see more freedom in quantum structures need some
empirical criteria, which would permit to detect the new geometries if
they exist. In case of classical state structures such criteria were
found by John Bell, in form of Bell inequalities expressing the
Boolean geometry of the state mixtures. Their breaking was the sign
that the ensembles are non-classical. The problems of quantum
ensembles, e.g., whether they indeed obey the Hilbert space geometry,
are significantly more involved. The initiative of our
colleagues~\cite{rf9} to describe them in terms of ``apophatic''
(forbidden) properties continues indeed the effort of John Bell on the
new theory level. Some interesting cases might be the ``cross
sections'' of $S$, resembling the ``constrained QM'' discussed
in~\cite{rf33}, and the projections (the collapsed $S$ caused by
deficiency of observables?). Simultaneously, the mathematical research
presented in~\cite{rf7,rf9} is an unexpected school of modesty for all
of us, who believed to understand so well the property of nice objects
called the ``density matrices''. Now it turns out that we did not even
know the properties of the simple qutrit! Needless to say, should any
of the ``forbidden properties'' be detected for any statistical
ensemble in some physical conditions, this will be the proof that the
theory is at the new conceptual level. Interesting, what about all
that will think the physicists of XXII~century?


\begin{thebibliography}{99}

\bibitem{rf1} G.~Birkhoff and J.~von~Neumann, Ann. of
  Math. \textbf{37}, 823--843 (1936).

\bibitem{rf2} J.~von~Neumann, \textit{Mathematical Foundations of
    Quantum Mechanics}, Princeton Unv. Press (1955).

\bibitem{rf3} D.~Finkelstein, Trans.~N.Y.~Acad. Sci. \textbf{25},
  621--637 (1962).

\bibitem{rf4} V.S.~Varadarajan, \textit{Geometry of Quantum Theory},
  Van Nostrand, vol~1 (1968), vol~2 (1970).

\bibitem{rf5} C.~Piron, Helv. Phys. Acta, \textbf{37}, 439--468
  (1964); Found. Phys. \textbf{2}, 287--314 (1972).

\bibitem{rf6} B.~Mielnik, Commun. Math. Phys. \textbf{15}, 1--45
  (1969).

\bibitem{rf7} I.~Bengtsson and K.~\.Zyczkowski. \textit{GEOMETRY OF
    QUANTUM STATES, An Introduction to Quantum Entanglement},
  Cambridge Univ. Press (2006).

\bibitem{rf8} B.~Mielnik, Commun. Math. Phys. \textbf{37}, 221--225
  (1974).

\bibitem{rf9} I.~Bengtsson, S.~Weis and K.~\.Zyczkowski,
  \textit{Geometry of the set of mixed quantum states: Apophatic
    approach}, preprint arXiv:1112.2347.

\bibitem{rf10} A.M.~Gleason, J. Math. Mech. \textbf{6}, 885--893
  (1957).

\bibitem{rf11} I.M.~Gel'fand and M.A.~Naimark, Math. Sbornik
  \textbf{12}, 197--213 (1943).

\bibitem{rf12} R.~Haag and D.~Kastler, J. Math. Phys. \textbf{5},
  848--861 (1964).

\bibitem{rf13} J.C.T.~Pool, Commun. Math. Phys. \textbf{9}, 118
  (1968); \textbf{9}, 212 (1968).

\bibitem{rf14} H.~Araki, Pacific J. Math \textbf{50}, 309--354 (1979.

\bibitem{rf15} R.~Haag, \textit{Local Quantum Physics, Fields,
    Particles, Algebras}, Springer-Verlag, 2-nd Edition (1996).

\bibitem{rf16} G.W.~Mackey, \textit{The mathematical foundations of
    quantum mechanics}, Benjamin, New York (1963).

\bibitem{rf17} H.~Primas, \textit{Chemistry, Quantum Mechanics and
    Reductionism, Perspectives in Theoretical Chemistry},
  Springer-Verlag, Berlin-Heidelberg (1983).

\bibitem{rf18} G.~Ludwig, Z.~Phys. \textbf{181}, 233--260 (1964).

\bibitem{rf19} E.B.~Davies and J.T.~Lewis,
  Commun. Math. Phys. \textbf{17}, 239--260 (1970).

\bibitem{rf20} R.~Penrose, Gen. Rel. Grav. \textbf{7}, 171--176
  (1976).

\bibitem{rf21} \textit{The Large, the Small and the Human Mind},
  Cambridge Univ. Press (1997).

\bibitem{rf22} T.W.B.~Kibble and S.~Randjbar-Daemi, J.~Phys.~A
  \textbf{13}, 141--148 (1980).

\bibitem{rf23} H.~Putnam, \textit{The Logic of Quantum Mechanics},
  Philosophical Papers, vol.~1, Cambridge Univ. Press (1975).

\bibitem{rf24} J.~Bell and B.~Hallet, Philosophy of Science
  \textbf{49}, 355--379 (1982).

\bibitem{rf25} I.~Bia{\l}ynicki-Birula and J.~Mycielski, Annals of
  Physics \textbf{100}, 62 (1976).

\bibitem{rf26} R.~Haag and U.~Bannier,
  Commun. Math. Phys. \textbf{60}, 1--6 (1978).

\bibitem{rf27} B.~Mielnik, Commun. Math. Phys. \textbf{101}, 323--339
  (1985).

\bibitem{rf28} N.~Gisin, Phys. Lett~A \textbf{143},1--2 (1989).

\bibitem{rf28a} C.~Simon, V. Buzek and N.~Gisin,
  Phys. Rev. Lett. \textbf{87}, 17 (2001).

\bibitem{rf29} D.J.~Fernandez and B.~Mielnik,
  J. Math. Phys. \textbf{35}, 2083 (1994).

\bibitem{rf30} F.~Delgado and B.~Mielnik, Phys. Lett.~A \textbf{249},
  369 (1998).

\bibitem{rf31} B.~Mielnik and O.~Rosas-Ortiz, J. Phys.~A \textbf{37},
  10007--10035 (2004).

\bibitem{rf32} B.~Mielnik and A.~Ramirez, Phys. Sci. \textbf{84},
  045008 (2011).

\bibitem{rf33} D.C.~Brody, A.C.T. Gustavsson and L.P.~Hugston,
  J.~Phys.~A \textbf{43} 082003 (2010).

\end{thebibliography}
\end{document}